\documentstyle[twoside]{article}

\catcode`\@=11
\long\def\@makefntext#1{
\protect\noindent \hbox to 3.2pt {\hskip-.9pt  
$^{{\eightrm\@thefnmark}}$\hfil}#1\hfill}		

\def\thefootnote{\fnsymbol{footnote}}
\def\@makefnmark{\hbox to 0pt{$^{\@thefnmark}$\hss}}	
	
\def\ps@myheadings{\let\@mkboth\@gobbletwo
\def\@oddhead{\hbox{}
\rightmark\hfil\eightrm\thepage}   
\def\@oddfoot{}\def\@evenhead{\eightrm\thepage\hfil
\leftmark\hbox{}}\def\@evenfoot{}
\def\sectionmark##1{}\def\subsectionmark##1{}}

\oddsidemargin=\evensidemargin
\renewcommand{\thefootnote}{\fnsymbol{footnote}}

\newcounter{sectionc}\newcounter{subsectionc}\newcounter{subsubsectionc}
\renewcommand{\section}[1] {\vspace{12pt}\addtocounter{sectionc}{1} 
\setcounter{subsectionc}{0}\setcounter{subsubsectionc}{0}\noindent 
	{\tenbf\thesectionc. #1}\par\vspace{5pt}}
\renewcommand{\subsection}[1] {\vspace{12pt}\addtocounter{subsectionc}{1} 
	\setcounter{subsubsectionc}{0}\noindent 
	{\bf\thesectionc.\thesubsectionc. {\kern1pt \bfit #1}}\par\vspace{5pt}}
\renewcommand{\subsubsection}[1] {\vspace{12pt}\addtocounter{subsubsectionc}{1}
	\noindent{\tenrm\thesectionc.\thesubsectionc.\thesubsubsectionc.
	{\kern1pt \tenit #1}}\par\vspace{5pt}}
\newcommand{\nonumsection}[1] {\vspace{12pt}\noindent{\tenbf #1}
	\par\vspace{5pt}}

\newcounter{appendixc}
\newcounter{subappendixc}[appendixc]
\newcounter{subsubappendixc}[subappendixc]
\renewcommand{\thesubappendixc}{\Alph{appendixc}.\arabic{subappendixc}}
\renewcommand{\thesubsubappendixc}
	{\Alph{appendixc}.\arabic{subappendixc}.\arabic{subsubappendixc}}

\renewcommand{\appendix}[1] {\vspace{12pt}
        \refstepcounter{appendixc}
        \setcounter{figure}{0}
        \setcounter{table}{0}
        \setcounter{lemma}{0}
        \setcounter{theorem}{0}
        \setcounter{corollary}{0}
        \setcounter{definition}{0}
        \setcounter{equation}{0}
        \renewcommand{\thefigure}{\Alph{appendixc}.\arabic{figure}}
        \renewcommand{\thetable}{\Alph{appendixc}.\arabic{table}}
        \renewcommand{\theappendixc}{\Alph{appendixc}}
        \renewcommand{\thelemma}{\Alph{appendixc}.\arabic{lemma}}
        \renewcommand{\thetheorem}{\Alph{appendixc}.\arabic{theorem}}
        \renewcommand{\thedefinition}{\Alph{appendixc}.\arabic{definition}}
        \renewcommand{\thecorollary}{\Alph{appendixc}.\arabic{corollary}}
        \renewcommand{\theequation}{\Alph{appendixc}.\arabic{equation}}
        \noindent{\tenbf Appendix \theappendixc #1}\par\vspace{5pt}}
\newcommand{\subappendix}[1] {\vspace{12pt}
        \refstepcounter{subappendixc}
        \noindent{\bf Appendix \thesubappendixc. {\kern1pt \bfit #1}}
	\par\vspace{5pt}}
\newcommand{\subsubappendix}[1] {\vspace{12pt}
        \refstepcounter{subsubappendixc}
        \noindent{\rm Appendix \thesubsubappendixc. {\kern1pt \tenit #1}}
	\par\vspace{5pt}}

\newcommand{\textlineskip}{\baselineskip=13pt}
\newcommand{\smalllineskip}{\baselineskip=10pt}

\def\eightcirc{
\begin{picture}(0,0)
\put(4.4,1.8){\circle{6.5}}
\end{picture}}
\def\eightcopyright{\eightcirc\kern2.7pt\hbox{\eightrm c}} 

\newcommand{\copyrightheading}[1]
	{\vspace*{-2.5cm}\smalllineskip{\flushleft
	{\footnotesize International Journal of Modern Physics A, #1}\\
	{\footnotesize $\eightcopyright$\, World Scientific Publishing
	 Company}\\
	 }}

\def\abstracts#1#2#3{{
	\centering{\begin{minipage}{4.5in}\baselineskip=10pt\footnotesize
	\parindent=0pt #1\par 
	\parindent=15pt #2\par
	\parindent=15pt #3
	\end{minipage}}\par}} 

\renewenvironment{thebibliography}[1]
	{\frenchspacing
	 \ninerm\baselineskip=11pt
	 \begin{list}{\arabic{enumi}.}
	{\usecounter{enumi}\setlength{\parsep}{0pt}
	 \setlength{\leftmargin 12.7pt}{\rightmargin 0pt} 
	 \setlength{\itemsep}{0pt} \settowidth
	{\labelwidth}{#1.}\sloppy}}{\end{list}}

\newcounter{itemlistc}
\newcounter{romanlistc}
\newcounter{alphlistc}
\newcounter{arabiclistc}

\newcommand{\fcaption}[1]{
        \refstepcounter{figure}
        \setbox\@tempboxa = \hbox{\footnotesize Fig.~\thefigure. #1}
        \ifdim \wd\@tempboxa > 5in
           {\begin{center}
        \parbox{5in}{\footnotesize\smalllineskip Fig.~\thefigure. #1}
            \end{center}}
        \else
             {\begin{center}
             {\footnotesize Fig.~\thefigure. #1}
              \end{center}}
        \fi}

\newcommand{\tcaption}[1]{
        \refstepcounter{table}
        \setbox\@tempboxa = \hbox{\footnotesize Table~\thetable. #1}
        \ifdim \wd\@tempboxa > 5in
           {\begin{center}
        \parbox{5in}{\footnotesize\smalllineskip Table~\thetable. #1}
            \end{center}}
        \else
             {\begin{center}
             {\footnotesize Table~\thetable. #1}
              \end{center}}
        \fi}

\def\@citex[#1]#2{\if@filesw\immediate\write\@auxout
	{\string\citation{#2}}\fi
\def\@citea{}\@cite{\@for\@citeb:=#2\do
	{\@citea\def\@citea{,}\@ifundefined
	{b@\@citeb}{{\bf ?}\@warning
	{Citation `\@citeb' on page \thepage \space undefined}}
	{\csname b@\@citeb\endcsname}}}{#1}}

\newif\if@cghi
\def\cite{\@cghitrue\@ifnextchar [{\@tempswatrue
	\@citex}{\@tempswafalse\@citex[]}}
\def\citelow{\@cghifalse\@ifnextchar [{\@tempswatrue
	\@citex}{\@tempswafalse\@citex[]}}
\def\@cite#1#2{{$\null^{#1}$\if@tempswa\typeout
	{IJCGA warning: optional citation argument 
	ignored: `#2'} \fi}}

\def\pmb#1{\setbox0=\hbox{#1}
	\kern-.025em\copy0\kern-\wd0
	\kern.05em\copy0\kern-\wd0
	\kern-.025em\raise.0433em\box0}

\def\fnm#1{$^{\mbox{\scriptsize #1}}$}
\def\fnt#1#2{\footnotetext{\kern-.3em
	{$^{\mbox{\scriptsize #1}}$}{#2}}}

\def\fpage#1{\begingroup
\voffset=.3in
\thispagestyle{empty}\begin{table}[b]\centerline{\footnotesize #1}
	\end{table}\endgroup}

\def\runninghead#1#2{\pagestyle{myheadings}
\markboth{{\protect\footnotesize\it{\quad #1}}\hfill}
{\hfill{\protect\footnotesize\it{#2\quad}}}}
\font\tenrm=cmr10
\font\tenit=cmti10 
\font\tenbf=cmbx10
\font\bfit=cmbxti10 at 10pt
\font\ninerm=cmr9

\font\eightrm=cmr8

\def\qed{\hbox{${\vcenter{\vbox{			
   \hrule height 0.4pt\hbox{\vrule width 0.4pt height 6pt
   \kern5pt\vrule width 0.4pt}\hrule height 0.4pt}}}$}}

\renewcommand{\thefootnote}{\fnsymbol{footnote}}	

\newcommand{\rf}[1]{(\ref{#1})}
\def\appendix#1{
  \addtocounter{section}{1}
  \setcounter{equation}{0}
  \renewcommand{\thesection}{\Alph{section}}
  \section*{Appendix \thesection\protect\indent \parbox[t]{11.15cm}
  {#1} }
  \addcontentsline{toc}{section}{Appendix \thesection\ \ \ #1}
  }

\def\al{{\lambda}}

 \def \const {{\rm const}}

\def\vm{{\mu}}
\def\vn{{\nu}}

\def\sfa{{\sf a}}

\def\dsfa{\dot{\sf a}}
\def\dsfb{\dot{\sf b}}

\def \bi{\bibitem}
\def \la {\label}

\def \ha {{1 \over 2}}
\def \ep{\epsilon}

\def \ov {\over}

\def\nline{\,\nabla\kern -0.7em\raise0.2ex\hbox{/}\,\,}
\def\yline{\,y\kern -0.47em /}
\def\aline{\,a\kern -0.49em /}
\def\parline{\,\partial\kern -0.55em /\,\,}

\def \t {\theta}
\def \s{\sigma}
\def \d {\partial}

\def\NPB#1(#2)#3{{\it Nucl. Phys.} {\bf B#1} (#2) #3}
\def\PRD#1(#2)#3{{\it Phys. Rev.} {\bf D#1} (#2) #3}
\def\PLB#1(#2)#3{{\it Phys. Lett.} {\bf B#1} (#2) #3}

\def\RMP#1(#2)#3{{\it Rev. Mod. Phys.} {\bf #1} (#2) #3}
\def\MPLA#1(#2)#3{{\it Mod. Phys. Lett.} {\bf A#1} (#2) #3}
\def\CQG#1(#2)#3{{\it Class. Quantum Grav.} {\bf #1} (#2) #3}
\def\AP#1(#2)#3{{\it Ann. Phys.} {\bf #1} (#2) #3}
\def\SJNP#1(#2)#3{{\it Sov. J. Nucl. Phys.} {\bf #1} (#2) #3}

\def \del{\partial}

\def\det{\hbox{det}}
\def\be{\begin{equation}}
\def\ee{\end{equation}}

\def \ci {\cite}
\def \g {\gamma}

\def \k {\kappa}

\def \m {\mu}
\def \n {\nu}

\def\x'{\mathaccent 19 x}
\def\y'{\mathaccent 19 y}
\def\n'{\mathaccent 19 n}
\def\u'{\mathaccent 19 u}

\def\X'{\mathaccent 19 X}
\def\Y'{\mathaccent 19 Y}
\def\Z'{\mathaccent 19 Z}

\def\et'{\mathaccent 19 \eta}
\def\th'{\mathaccent 19 \theta}
\def\lam'{\mathaccent 19 \lambda}
\def\varet'{\mathaccent 19 \vartheta}
\def\rh'{\mathaccent 19 \rho}
\def\ph'{\mathaccent 19 \phi}
\def\xb'{\mathaccent 19 {\bar{x}}}
\def \adss {$AdS_5 \times S^5$\ }
\def \N {{\cal N}}
\def \lc {light-cone\ }
\def \ta { \tau}
\def \s { \sigma }

\def \sg {\sqrt {g }}
\def \te {\theta}

\def \xp {x^+}

\def \p {\phi}
\def \vt {\theta}
\def \bx {\bar x} \def \a { \alpha}

\def \ri {{\rm i}}

\def \diag {{\rm diag}}

\def \al {\perp}
\def \PP {{\cal P}} 

\begin{document}
 
\runninghead{Superstrings in AdS ...
} 
{Superstrings in AdS ...}

\normalsize\textlineskip
\thispagestyle{empty}
\setcounter{page}{1}

\copyrightheading{}			

\vspace*{0.88truein}

\fpage{1}
\centerline{\bf SUPERSTRINGS IN ADS IN  LIGHT CONE GAUGE}

\vspace*{0.37truein}
\centerline{\footnotesize A.A. TSEYTLIN\footnote{ Also at Imperial
College, London and Lebedev Institute, Moscow}}
\vspace*{0.015truein}
\centerline{\footnotesize\it Physics  Department, The Ohio State
University}
\baselineskip=10pt
\centerline{\footnotesize\it  Columbus, OH43210-1106,
 USA\footnote{E-mail: tseytlin.1@osu.edu }}


\vspace*{0.21truein}
\abstracts{We discuss  light-cone gauge description of 
type IIB Green-Schwarz superstring in 
 $AdS_5 \times S^5$ with a hope to make progress towards
 understanding spectrum of this theory.
  As in flat space, fixing  \lc gauge 
  consists of
   two steps: (i) fixing kappa symmetry in such a way that the
 fermionic part of the
 action does not depend on $x^-$; (ii)  fixing  2-d
 reparametrizations by $x^+=\tau$ and a condition on 2-d metric. 
 In curved AdS space the latter cannot be 
 the standard conformal gauge and breaks manifest
 2-d Lorentz invariance.  It is natural, therefore,  to  work  in 
 phase-space framework, imposing
 the  GGRT \lc gauge conditions   $x^+=\tau, \ P^+ =$const.
 We obtain the resulting \lc superstring  Hamiltonian.
 }{}{}


\vspace*{1pt}\textlineskip	
\section{Introduction}	
\vspace*{-1.5pt}
\noindent
\textheight=7.8truein
\setcounter{footnote}{0}
\renewcommand{\thefootnote}{\alph{footnote}}

\noindent
One lesson of   developments in the last 4 years is 
 that  string theory duals of 
($\N=4,2,1,0$) supersymmetric  gauge theories 
 should be  fermionic strings in  4+1+more  dimensions 
 propagating in curved space with Ramond-Ramond backgrounds
 (see \ci{Pol,mald} and, e.g., \ci{KSMN,KLT,Krev}
 for reviews and references).
 It is important to  make progress in understanding  properties
 of such string theories.
 
 The basic most symmetric example   is  
   (large $N$) $\cal N$=4 Super Yang-Mills 
 theory dual 
to (weakly coupled) type IIB superstring theory in \adss space 
with Ramond-Ramond flux \ci{mald}. 
Though this background has a lot of symmetry, solving
the corresponding string theory appears   to be a 
complicated problem. The commonly used procedure, in the known 
exactly solvable cases,
 is to start with
 a  string  action, solve the classical string equations, 
 then quantize the theory, find the string spectrum, 
 vertex operators, scattering amplitudes, etc.
Even the first steps in this program are nontrivial in the 
\adss  case.

In contrast to other known cases, here  fermionic 
degrees of freedom play  crucial role and cannot be ignored from
the start.
 The bosonic part of the string sigma model is the sum of the 
two  symmetric space chiral models --
$SO(2,4)/SO(1,5)$ and $SO(6)/SO(5)$. These are not conformally
invariant ($R_{mn}\not=0$), and it is the fermionic R-R couplings 
($\bar \theta \gamma^{...}  \theta \del x \del x F_{5...}$) 
that ``glue" the $AdS_5$ and $S^5$ bosons together 
 ensuring  the vanishing of the 2-d beta-function
($R_{mn} -( F^2_5)_{ mn} =0$). 

The presence of a curved R-R background indicates that one should
use 
the manifestly supersymmetric Green-Schwarz \ci{GS} description 
of superstring. Finding an explicit expression
for  curved space GS  action  $S(x,\t)$    is 
difficult in general
 (one needs to know the component
expansion  of the  background supergravity
superfields which enter the formal expression 
\ci{howe} for the action). In the present \adss case, this technical problem
 has a nice geometrical
solution \ci{MT1} based on viewing string as moving on  the supercoset
$PSU(2,2|4)/[SO(1,4) \times SO(5)]$  which replaces the flat
superspace
(10-d super Poincare)/$SO(1,9)$ in  original GS construction. 

The resulting 
 action, though  explicitly  known \ci{MT1,KRR,MT2}, 
 looks  highly
 nonlinear
 containing  terms of many  orders in $\t$. 
 This is, however, an illusion of complexity:
 the fermionic part of the action simplifies dramatically 
 in  proper $\k$-symmetry gauges -- it   becomes  quadratic
 and quartic in $\t$ only \ci{Pes,KR,KT}. Its 
 structure is similar to that of the flat space   GS action 
 in a covariant $\k$-symmetry gauge  which  is also 
 quartic in fermions.\fnm{a}\fnt{a}{For example, in $\t^1=\t^2$ gauge 
 the flat space type IIB GS action has the following 
 structure:  
 $S \sim \int [(\del x)^2 + \del x \bar \t \g \del \t + (\bar \t \del
 \t)^2]$.}

 If one interprets the   \adss supergroup  $PSU(2,2|4)$ as 
 the   $\N=4$ 
 superconformal group in 4 dimensions, it is natural to split 
 the  fermionic generators into 4 standard supergenerators $Q_i$
 and  4 special conformal supergenerators $S_i$ (we suppress the  
 4-d spinor indices). The associated 
 superstring coordinates will be denoted as 
 $\theta_i$ and $\eta_i$ (we use fermionic parametrization of
 \ci{MT3,MTT}).
 The 4-d Lorentz covariant $\k$-symmetry 
  ``S-gauge" \ci{MT3} (which 
 leads to an action equivalent to the one in 
 \ci{Pes,KR}) 
 corresponds to setting all $\eta_i$ to zero. The resulting 
 superstring Lagrangian  written in  ``4+6" parametrization 
 of \adss  where ($R=1$)   
 \be
 ds^2 = Y^2 dx^a dx^a + Y^{-2} dY^MdY^M \ , \la{mmm}
 \ee
 or, equivalently, ($a=0,...,3, \ 
 M=1,...,6$, $Y^M= e^\p u^M$)
  \be 
 ds^2= \ e^{2 \p} dx^a dx^a  + d\p^2 +   du^M du^M\ , \ \ \  \ \ \ \ 
   u^M u^M =1 \ , 
 \la{gtg}
 \ee
 has the following simple structure \ci{MT3}\fnm{b}\fnt{b}{The actions in 
 \ci{Pes,KR} have  isomorphic form, corresponding to a specific
 choice
 of the 10-d Dirac matrix representation.
 We use  ``chiral" representations for the 4-d and 6-d Dirac matrices,
 $\g^a =\left(\begin{array}{cc}
 0   & \s^a
 \\
\bar  \s^a & 0
 \end{array}
 \right)$, \ 
 $ \gamma^M
=\left(\begin{array}{cc}
 0   & \rho^{Mij}
 \\
 \rho^{M}_{ij} & 0
 \end{array}
 \right)\,, $  with 
 $(\rho^M)^{ij}\equiv  - (\rho_{ij}^{M})^*$, \ $i,j=1,2,3,4$.
 $\sfa,\dsfb=1,2$ are   the  $sl(2,C)$ 
 (4-d spinor) indices and 
 $\theta_{\sfa i}^\dagger = -\theta_{\dsfa}^i$, 
 $\theta^{\sfa}_i{}^\dagger = \theta^{\dsfa i}$. 
 The 10-d spinors are split as $(\t^i,\t_i)$.}
 $$
 L= - \frac{1}{2}\sqrt g 
 \Bigl( Y^2 [\del_\vm x^a
  - ({\rm i} \theta_{\sfa i} 
  \sigma^{a\sfa\dsfb}\del_\vm \theta_{\dsfb}^i 
     + h.c.)]^2 
 + Y^{-2}\del^\vm Y^M \del_\vm Y^M \Bigr) $$
 \be 
  - \bigl( {\rm i} \epsilon^{\mu\nu} \del_\mu Y^M 
   \theta^{\dsfa i} \rho^M_{ij} \del_\nu  \theta_{\dsfa}^j +
   h.c.\bigr)  \  . \la{bss}
   \ee
This  covariant $\k$-symmetry gauge fixed  action is 
 well-defined and useful for developing perturbation 
theory near classical ``long'' string  configurations 
ending at the boundary of $AdS_5$ \ci{KT,Forste,DGT}, e.g., 
 the ones appearing in  Wilson loop computations
\ci{malda}. However, the  kinetic term of the fermions 
$ \sim  \del x \theta  \del \theta $
 is degenerate for ``short'' strings \ci{KT}, and thus 
 this action is not directly applicable 
  for  computing  the spectrum of ``short" 
  closed string in the bulk 
  of \adss.
  
In order to  avoid  this degeneracy problem,
 it is natural to try to follow 
the same approach which worked remarkably well 
 in flat space \ci{GSlc,GS}:  use  \lc gauge 
 (for an alternative covariant approach to quantization 
 of GS action  see \ci{berk}).
 In flat space superstring \lc gauge  fixing consists of the two 
 steps:
(i) fermionic \lc gauge choice,
i.e. fixing the $\k$-symmetry by  $\g^+ \theta^I=0$;
(ii) bosonic \lc gauge choice, i.e.
using   the conformal gauge\fnm{c}\fnt{c}{We  use Minkowski
signature 2-d world sheet  metric $g_{\vm\vn}$ with
$g\equiv - \det g_{\vm\vn}$.}
$\sqrt {g}  g^{\vm\vn} =\eta^{\vm\vn}$
and fixing  the  residual conformal diffeomorphism symmetry
by $x^+(\ta,\s)  =  p^+ \tau$.
Fixing the  fermionic \lc gauge
  already produces a substantial
simplification of the flat-space
GS action: it   becomes  quadratic in $\theta$.
Similarly, in \adss the first task  should be to  
  find a light-cone 
  $\k$-symmetry gauge in  which the fermion kinetic term 
  becomes $  \del x^+ \theta  \del \theta $, i.e.
    involves only one
  combination -- $x^+$ -- of 4-d coordinates, so that  
 the non-degeneracy of the kinetic term
for the  fermions  will not
 depend on a choice of a specific  string background
  in transverse directions.
  To simplify the fermion kinetic term further 
  one should then try to   choose the  light-cone 
  bosonic gauge $x^+=\tau$. 
  
Fixing the fermionic \lc gauge was discused in detail 
in \ci{MT3},\fnm{d}\fnt{d}{Some 
  previous work in
 this direction using different approach
was reported in \ci{pes}.}
and the fixing the bosonic part of \lc gauge 
and derivation of the resulting \lc Hamiltonian 
was described  in \ci{MTT}. Our aim below is to review 
these results.


The metric \rf{gtg}  corresponds to 
the $AdS$ space  in the  Poincar\'e  coordinate patch.
One needs to use the Poincar\'e coordinates to have a null
 isometry in the bulk  and  at the boundary (the boundary
 should have  $R^{1,3}$ topology).
The      AdS/CFT duality suggests that 
since  the boundary SYM theory 
in $R^{1,3}$  has a well-defined light-cone gauge 
description \ci{brink},
 it should be possible to fix 
 some analog of a \lc gauge  for the dual string theory as well. 
 
 The SYM  theory does not admit a
 manifestly $\N=4$  supersymmetric  Lorentz-covariant description,
 but has a  simple superspace description in the
 \lc gauge $A^+=0$
 \ci{brink}. It is  based
 on a single  chiral superfield
 $\Phi(x, \te) = A(x) + \theta^i \psi_i(x) + ...$   where
 $A= A_1 + \ri A_2$ represents the  transverse
 components of the gauge field and $\psi_i$  its fermionic partner
 which transforms under the fundamental representation of R-symmetry
 group
 $SU(4)$.
 In addition to the  standard \lc  supersymmetry
 (shifts of  $\theta$),
 the \lc superspace SYM  action $S[\Phi]$
  has  also a non-linear superconformal symmetry.
This suggests that it may be possible   to formulate the bulk
string theory  in a way which is  naturally related to the \lc  form
of the boundary SYM theory. In particular,  it may be useful
to  split the   fermionic string   coordinates
into  the   two  parts  with  manifest $SU(4)\simeq SO(6)$
transformation properties
which  will be the counterparts of the linearly realized
Poincar\'e supersymmetry  supercharges $Q_i$ 
and  the  nonlinearly realized conformal supersymmetry 
 supercharges $S_i$ 
of the  SYM  theory.

 As was shown in \ci{met1,met2,met3},
field theories  in AdS space  (in particular, type  IIB supergravity)
admit a   simple  \lc description. There exists
a \lc   Hamiltonian  for a superparticle in \adss
which may be  used to formulate AdS/CFT correspondence between 
chiral primary states of SYM theory and supergravity states of the
bulk theory in the \lc gauge.
This  indicates  that the full superstring theory 
in \adss should also
have a  kind of  \lc gauge  formulation, which may  be useful
in the context of   the AdS/CFT correspondence.
The string \lc  Hamiltonian should reduce to the flat space one in
the $R\to \infty$ limit  and to the superparticle Hamiltonian
\ci{met3} in the zero slope limit.

This is the motivation  behind our \lc gauge 
fixing program \ci{MT3,MTT}, in spite of
the fact that  there is no
 globally well-defined null Killing  vector  in $AdS$ space
 (its  norm proportional to $e^{2\phi}$
 vanishes at the horizon $\phi=-\infty$ of the  Poincar\'e 
 patch).
 We use a formal approach,  
assuming that the  degeneracy of the 
\lc  description  reflected in the
$e^{2\p} \to 0$ singularity  of the resulting \lc gauge fixed action 
should have some    physical resolution, e.g., 
 the form of the 
\lc  Hamiltonian  may  suggest how  the wave functions 
 should be defined in this region.

\section{Superstring Action in \lc kappa symmetry gauge}
\noindent

In  \ci{MT3} it was shown  how 
the  light-cone type $\k$-symmetry gauge can be fixed
in the original action of \ci{MT1} so that 
 the resulting action   has indeed the required form:
 only $\del x^+$ appears in the fermionic kinetic term.
In contrast to the Lorentz covariant ``S-gauge" where 16 fermions 
 $\eta_i$ are set equal to zero, the light-cone 
 gauge  used in  \ci{MT3} corresponds to setting to zero 
 ``half" of the 16 $\theta_i$ and ``half" of the 16 $\eta_i$
 (``half" is defined with respect to $SO(1,1)$ rotations in the 
 light-cone directions). The remaining fermions
(which will be again   denoted  by  $\theta_i, \eta_i$)
from  now on are  simply 4+4  complex anticommuting variables 
 carrying no extra  Lorentz spinor  indices.
 For comparison,  the flat space  GS action in the 
 light-cone gauge ($ \gamma^+ \theta=0$) 
  \ci{GSlc}, written in a  similar parametrization 
 of the 16 fermionic
 coordinates, has the following structure:
  \be
{\cal L}
=- \frac{1}{2}\sqrt{g} (\partial_\vm x)^2 - \Bigl[
 \frac{{\rm i}}{2}\sqrt{g} \partial^\vm x^+(
\theta\partial_\vm \theta
+\eta\partial_\vm\eta)  - \epsilon^{\vm\vn}
\partial_\vm x^+ \eta\partial_\vn\theta+h.c.
\Bigl]\ .\ee Note that $\theta$'s and $\eta$'s
enter diagonally in the kinetic term, but 
 they are mixed in the WZ term. 
This form of the  original  GS Lagrangian  is the 
flat space limit
 of  the  \lc  \adss  Lagrangian   of \ci{MT3} (cf. \rf{bss})
$$
{\cal L} =
-\sqrt{g}\Bigl[
Y^2(\partial^\vm x^+ \partial_\vm x ^-
+ \partial^\vm x\partial_\vm\bar{x})
+\frac{1}{2} Y^{-2} ( \del_\mu Y^M +  
{\rm i}\eta\rho^{MN}\eta  Y^N Y^2 \del_\mu x^+ )^2      \Bigl]
$$
$$ - \ \frac{{\rm i}}{2} \sqrt{g}g^{\vm\vn}
Y^2\partial_\vm x^+
\Bigl[\theta^i\partial_\vn \theta_i
+\theta_i\partial_\vn \theta^i
+\eta^i\partial_\vn \eta_i
+\eta_i\partial_\vn \eta^i 
+{\rm i}  Y^2\partial_\vn x^+(\eta^2)^2\Bigr] 
$$ \be
+ \ \epsilon^{\vm\vn}
|Y|\partial_\vm x^+ \Bigl[  \eta^i \rho_{ij}^M Y^M
(\partial_\vn\theta^j-{\rm i}\sqrt{2}|Y| \eta^j
\partial_\vn x)+h.c.\Bigl] \ . \label{actki}
\ee
We decomposed $x^a$ into the light-cone and  2 complex
coordinates:
$(x^+,x^-,x, \bx)$, \ \  $
x^\pm= \frac{1}{\sqrt{2}}(x^3\pm x^0)\,; $ $
x,\bx = { 1 \ov \sqrt 2} (x^1 \pm  {\rm i} x^2)$.\fnm{e}\fnt{e}{We use the following notation:
4-d indices are $a,b=0,1,2,3$; $SO(6)$ indices are 
 $ M,N=1,...,6$; $SU(4)$ indices are $i,j=1,2,3,4$;
2-d indices are 
$\mu,\nu=0$.  
The matrices $\rho^M$ (off-diagonal blocks of 6-d Dirac matrices
in chiral representation) satisfy:
$\rho_{ij}^M =- \rho_{ji}^M\,,$ $ 
   (\rho^M)^{il}\rho_{lj}^N + (\rho^N)^{il}\rho_{lj}^M
 =2\delta^{MN}\delta_j^i\,, $ $
  (\rho^M)^{ij}\equiv  - (\rho_{ij}^{M})^* ,$
  and $\rho^{MN} \equiv \rho^{[M} \rho^{\dagger N]} $, i.e. 
  $\rho^{MNi}_{\ \ \ j} =\frac{1}{2}(\rho^M)^{il}\rho_{lj}^N
-(M\leftrightarrow N)\,.$ 
Also, $ \theta_i^\dagger =\theta^i\,,$ $
\eta_i^\dagger = \eta^i\ , $
$
\theta^2 \equiv \theta^i\theta_i\,, \ 
\eta^2 \equiv \eta^i\eta_i\,.$ } 
Written in terms of the radial direction $\p$ of $AdS_5$ 
and unit 6-d  vector $u^M$  parametrizing $S^5$ this
Lagrangian  becomes ($Y^M= e^\p u^M$, $|Y|= e^\p$) \ci{MT3}
$$
{\cal L} =
-\sqrt{g}\Bigl[
e^{2 \p} (\partial^\vm x^+ \partial_\vm x ^-
+ \partial^\vm x\partial_\vm\bar{x}) + \frac{1}{2} \del^\mu \p
\del_\mu \p 
+\frac{1}{2}  ( \del_\mu u^M +  
{\rm i}\eta\rho^{MN}\eta  u^N  e^{2\p}  \del_\mu x^+ )^2      \Bigl]
$$
$$ - \ \frac{{\rm i}}{2} \sqrt{g}g^{\vm\vn}
 e^{2\p} \partial_\vm x^+
\Bigl[\theta^i\partial_\vn \theta_i
+\theta_i\partial_\vn \theta^i
+\eta^i\partial_\vn \eta_i
+\eta_i\partial_\vn \eta^i 
+{\rm i}  e^{2\p} \partial_\vn x^+(\eta^2)^2\Bigr] 
$$ \be
+ \ \epsilon^{\vm\vn}
e^{2\p} \partial_\vm x^+\Bigl[  \eta^i \rho_{ij}^M u^M
(\partial_\vn\theta^j-{\rm i}\sqrt{2}e^{\p}  \eta^j
\partial_\vn x)+h.c.\Bigl] \ . \label{tki}
\ee
This action has several important properties:

(a) It contains  $x^-$  
only in the bosonic part 
and only linearly (in $\del x^+ \del x^-$ term), 
and the fermion kinetic terms are multiplied by 
the derivative of $x^+$ 
only. 
It is the crucial property of this \lc $\k$-symmetry gauge
fixed form of the action  that the fermion  kinetic
term involves  the derivative of
only  {\it one} space-time direction
-- $x^+$, i.e. its (non)degeneracy
does not depend on  transverse
string  profile. One expects, therefore, that
after  one fixes  the bosonic \lc gauge, 
 it should this action should be 
  a well defined
starting point for quantizing the theory in the 
``short'' string sector.

(b) The  fermionic  $\k$-symmetry
\lc gauge  we used 
 reduces the 32 fermionic coordinates $\theta^I_\a$ (two left
Majorana-Weyl 10-d spinors)
to 16 physical Grassmann variables:
``linear"  $\te^i$  and   ``nonlinear" $\eta^i$
and their hermitian conjugates $\te_i$  and $\eta_i$,
which  transform
in  fundamental representations of $SU(4)$.
The superconformal algebra $psu(2,2|4)$  implies  that
these   are  related to  the  Poincar\'e
and  the  conformal
supersymmetry   in the \lc gauge description of the
    boundary theory. The action and symmetry generators
     have simple (quadratic)
 dependence on
$\te^i$, 
but complicated (quartic)
dependence on $\eta^i$. 
The action is symmetric under shifting $\theta\rightarrow
\theta+\epsilon$
  (supplemented by a shift  of $x^-$).
 It is this symmetry that is
responsible for the fact that
the theory is linear in $\theta$, i.e.
that there is no quartic terms in $\theta$.

(c) The fact that the action  has  only  quadratic and quartic
fermionic terms  has to do with special symmetries
of the \adss background (covariantly
constant curvature and 5-form  field strength).
The presence of the  $\eta^4$ term   reflects
the curvature of the background.\fnm{f}\fnt{f}{Note that
the  light-cone gauge
GS action in a curved space of the form $R^{1,1} \times M^8$
with generic  NS-NS and R-R backgrounds  \ci{FRT}
(reconstructed
from the \lc flat space  GS vertex operators)
contains,  in general,  higher than quartic fermionic terms,
multiplied by higher derivatives of the background fields.
This     \lc  GS
action has  quartic fermionic term \ci{HW,FRT}
involving the curvature tensor \
$R_{....} \d \xp  \d \xp  (\bar \theta\g^{-..} \vt )(\bar \theta 
\g^{-..} \theta)
\sim
R_{....} (p^+)^2  (\bar \theta \g^{-..} \theta)  ( \bar \theta
\g^{-..} \theta) $ which is similar to the one present in the
NSR string action
(i.e. in  the standard 2-d supersymmetric sigma model).}
As follows from the discussion in \ci{MT1},
the  `extra' $O(\eta^2)$ terms  
should have the interpretation of the
couplings to the R-R  5-form  background.

(d) The  \adss superstring action in general 
 depends on  two  parameters:
the scale   (equal radii)  $R$ of $AdS_5\times S^5$ and the inverse
string tension
or $\alpha'$.
Restoring the dependence on    $R$
 one finds that in the flat space limit    $R\to \infty$
 the quartic fermionic term  goes away, while
  the kinetic term 
 reduces to the  flat space
 \lc GS action  after representing  each of the
 two $SO(8)$ spinors in terms of the two  $SU(4)$ spinors.

\section{Fixing the bosonic \lc  gauge}
\noindent

As a next step  to quantization of the theory
one would like, as in the flat   case,
 to eliminate the $\del x^+ $-factors from the fermion kinetic
 terms  in \rf{tki}.
 In flat space this  was  possible by choosing the  bosonic \lc
 gauge.    In the  Polyakov  formulation 
 this may be  done by fixing the conformal gauge
   $
   \sg g^{\vm\vn}=\eta^{\mu\nu} ,$
     and then noting
   that since the resulting equation
   $\del^2 x^+=0$ has the general solution
   $x^+ (\tau,\s) = f( \tau-\s) + h( \tau+ \s)$, 
   one can fix the residual conformal diffeomorphism symmetry
   on the plane by choosing $x^+(\tau,\sigma) = p^+ \tau$.

Let us  first not make any explicit gauge choice
 and consider the superstring path integral
  assuming   that there is no sources for $x^-$.
The linear dependence of the action \rf{actki},\rf{tki}
on $x^-$  allows  us  to  integrate over    $x^-$
explicitly: we get $\delta$-function constraint  imposing the equation of
 motion    $\del_\mu ( \sqrt g g^{\mu\nu} e^{2\p} \del_\nu x^+)=0$ 
 for $x^+$, which
  is formally solved by    setting
\be
\sg g^{\vm\vn} e^{2 \p} \del_\vn x^+ = \ep^{\vm\vn} \del_\vn  f
\ , \ \ \ {\rm i.e.} \ \ \  \ \ \ \ 
 e^{2 \p} \del_\mu x^+ = 
  g_{\mu\nu }{ \ep^{\nu\lambda} \ov \sg } \del_\lambda 
  f\ .  
\la{solv}
\ee
where $f(\tau,\s)$ is an { arbitrary}  function.
Since our action \rf{tki}  depends  on  $x^+$
only through $e^{2 \p} \del x^+ $,
we are then   able to integrate over $x^+$ as well,
eliminating it
in favor  of the function  $f$.
The resulting fermion kinetic term is  then
non-degenerate (for a properly chosen  $f$), and may  be
interpreted as  an  action of 2-d fermions in curved 2-d geometry
 determined by $f$ and $g_{\mu\nu}$ (cf. \ci{DGT,MT3} and refs. there).
 
To proceed further, one would like to fix 2-d diffeomorphisms
by a condition on $g_{\mu\nu}$
 and  a condition on $f$ corresponding to the usual 
 \lc gauge $x^+=\tau$.
An important  observation \ci{rudd,pol,MT3} is that 
in the case of the  $AdS$ type curved spaces, 
the bosonic \lc gauge $x^+=\tau$   
can not be combined with the standard conformal gauge
$\sqrt g g^{\m\nu}= \eta^{\mu\nu}$ (imposing the conformal gauge 
one in general is unable  to  solve 
the equation for $x^+ $ by $x^+=\tau$). Instead,   
one needs to impose a condition
on $g_{\m\nu}$ that breaks the manifest  2-d Lorentz symmetry
and leads to a rather  non-standard  string action,
with all terms coupled to the radial function $\phi $
of $AdS_5$ space.

 A consistent gauge choice  is \ci{MT3} 
  \be
 e^{2\phi}\sqrt g g^{00}=-1\ , \   \ \ \ \ \ \ \ \ \  
 \ x^+ =  p^+ \tau  \ .
 \la{our}
 \ee
 Indeed,  the  equation for $x^+$
 is then satisfied identically. This choice is equivalent to 
 $f=\s$ in \rf{solv}.
A closely related  
alternative,  originally suggested by Polyakov \ci{pol},
is a  modification of the conformal gauge 
$\sqrt g g^{ab} = \diag(- e^{-2\p}, e^{2\p})$ 
which is also  consistent with  the \lc gauge condition 
$x^+=\tau$.
   
 The resulting action has $AdS_5$ radial direction $\p$
 factors coupled differently to $\del_0$ and $\del_1$
 derivative terms, and   the $S^5$ part of
  the action is also coupled to $\p$. 
The  absence  of manifest 2-d Lorentz symmetry  suggests that 
it is natural to use the {\it phase space}  formulation   of 
the \lc gauge fixed  theory. 
The coordinate space Polyakov 
   approach based on \rf{our} is equivalent
 to the phase space approach 
   based on  fixing the
 diffeomorphisms by 
   $x^+ = p^+ \tau, \ P^+  =$const \ci{MTT}. 
In general, the original GGRT 
phase space approach \ci{ggrt}  to \lc gauge fixing 
is directly applicable in the present curved space case.


To illustrate the derivation of phase space Lagrangian  
let us consider first the  example of a
classical  bosonic string in $AdS_5$
($2 \pi \a'=1, \ R=1$) 
\be
{\cal L} = -    h^{\vm\vn} \left(  
\partial_\vm x^+ \partial_\vn x^-
+ \partial_\vm x\partial_\vn\bar{x}
 + \ha e^{-2\p}\del_\vm \p\del_\vn \p\right) \ , 
 \qquad\ \ 
 h^{\vm\vn} \equiv \sqrt g g^{\vm\vn} e^{2\p}  \ . 
 \la{lll}
 \ee
Introducing the momenta $\PP^a$ for the \lc coordinates $(x^+,x^-)$
and the two transverse coordinates $(x,\bar x)$ and the momentum $\pi$ for 
the radial direction $\p$ we get 
\begin{eqnarray} 
{\cal L} = && \dot x_\al  \PP_\al   + \dot \p \pi + \dot x^+ \PP^-  +
\dot x^- \PP^+   
\nonumber\\ 
&+& \  { 1 \ov 2 h^{00}} 
\bigg[(  \PP^2_\al + 
2\PP^+\PP^-) + e^{4\p}   (  \x'^2_\al  +  2\x'^+ \x'^-) + e^{2\p} 
(\pi^2 + \ph'^2 ) 
\bigg]
\la{logj} \nonumber \\ &+& \  { h^{01} \ov  h^{00} }
( \x'_\al \PP_\al 
+ \ph'\pi  + \x'^+ \PP^-  + \x'^- \PP^+) \ , 
\end{eqnarray} 
where $ 1/h^{00}$ and  ${ h^{01}/h^{00} }$ play the role of the
Lagrange
multipliers imposing two constraints. 
Choosing the \lc gauge  $
  x^+ = \tau\ , \  \PP^+  = p^+  = \const $  and 
integrating over $\PP^-$ we get the relation 
\be
h^{00}    = - p^+  \ ,  \la{guu}
\ee
which is equivalent (up to a rescaling) to the 
condition on the metric in \rf{our}.
The expression for $\PP^-$ follows from the $1/h^{00}$
constraint after using  the $h^{01}/h^{00}$  constraint.
The resulting \lc gauge Hamiltonian density is  \ci{MTT}  
\be
{\cal H} =-\PP^- = {1 \ov 2 p^+ } 
  \bigg[  \PP^2_\al  + e^{4\p}  \x'^2_\al 
 +  e^{2\p}(\pi^2 +  \ph'^2)  \bigg] \ . 
\la{hamm}
\ee
As usual, the coordinate $x^-$ does
not appear in the Hamiltonian, but is determined 
 from  the reparametrization
constraint
\be
p^+\x'^-  + \PP_\al\x'_\al   + \ph'\pi =0 
 \ .  \la{coons}
\ee
For  closed  string, the integral of this constraint over
$\sigma$ constrains the state space to the subspace
invariant under $\sigma$ translations. 

For comparison, for string in $AdS_3$ described by 
$SL(2,R)$ WZW model (in the standard Gauss
parametrization) \be 
{\cal L}
=-\sqrt g g^{\vm\vn}(e^{2\phi}\partial_\vm x^+\partial_\vn  x^-
 +\frac{1}{2}\partial_\vm \phi \partial_\vn \phi)
+\epsilon^{\vm\vn}e^{2\phi}\partial_\vm x^+\partial_\vn x^-
\ , \ee
 one finds 
\be {\cal H}  = 
\frac{e^{2\phi}}{2p^+}(\pi + \ph')^2 \ , \ \ \ \ \ \ \ \ \ \ \ 
 p^+\x'^- +  \pi\ph'=0 \ . 
\ee

\pagebreak
\section{Light cone superstring Hamiltonian }
\noindent

Repeating the same procedure for the superstring action
in \rf{tki} we get \ci{MTT} \fnm{g}\fnt{g}{We  
explicitly introduce momenta  only for the  bosons.  
The odd (fermionic) 
part of the  phase space  may be viewed as 
 represented by 
 $\theta^i$, $\eta^i$ considered as fermionic 
coordinates and $\theta_i,$ $\eta_i$ considered as 
fermionic momenta.} 
$$
{\cal L} 
= 
\PP_\al\dot{x}_\al +  \pi\dot{\phi} 
+ \PP_M \dot{u}^M
+\frac{{\rm i}}{2}p^+(\theta^i \dot{\theta}_i
+\eta^i\dot{\eta}_i+\theta_i
\dot{\theta}^i+\eta_i\dot{\eta}^i) + \PP^- 
$$
\be - \ 
\frac{h^{01}}{p^+}\Bigl[p^+\x'^-  + \PP_\al\x'_\al + \pi\ph'
+ \PP_M \u'^M
+\frac{{\rm i}}{2}p^+(\theta^i\th'_i+\eta^i\et'_i
+\theta_i\th'^i+\eta_i\et'^i)\Bigr] \  , \label{lag6}
\ee
where  the \lc Hamiltonian density is (${\cal H} =-\PP^-$)
$$
{\cal H} = 
\frac{1}{2p^+}\Bigl(\PP_\al^2+e^{4\phi}\x'_\al^2 $$ $$ 
+\ e^{2\phi}[\pi^2+\ph'^2
+(l^i{}_j)^{2} + \u'^M \u'^M
+p^{+2}(\eta^2)^2 + 4p^+\eta_i l^i{}_j \eta^j] 
\Bigr) $$ 
\be
-\ 
e^{2\phi} u^M \Bigl[ \eta^i \rho^M_{ij} 
(\th'^j - {\rm i}\sqrt{2}e^\phi \eta^j\x')
\    +   \  \eta_i \rho^{Mij} 
(\th'_j + {\rm i}\sqrt{2}e^\phi \eta_j\xb')\Bigr] \, ,\la{dur}
\ee
and $h^{01}$ imposes the reparametrization constraint. 
Here  $\PP^M$ is the momentum  corresponding to the unit 6-vector
$u^M$, \ 
$u^M \PP^M =0$,  and   $
l^i{}_j \equiv  { {\rm i}  \over 2} (\rho^{MN})^i{}_j u^M \PP^N ,$
so that 
$(l^i{}_j)^{2}\equiv l^i{}_j l^j{}_i
=\PP^M\PP^M$. In
 the particle theory limit $
l^i{}_j $ has the interpretation of the $SU(4)$ angular momentum
operator \ci{met3} and \rf{dur} reduces to the 
(classical limit of) \lc 
Hamiltonian for a superparticle in \adss space \ci{met3}
\be
{\cal H} 
= \frac{1}{2p^+}\Bigl[\PP_\al^2
+e^{\phi}\pi e^\phi \pi +e^{2\phi} (X- \frac{1}{4}) \Bigr] \  , 
\ee
\be
X\equiv 
(l^i{}_j)^{2} +(p^+\eta^2-2)^2 + 4p^+\eta_i l^i{}_j \eta^j \ . 
\ee
The analogous expressions  for the string Lagrangian and Hamiltonian
written in terms  of 6 Cartesian coordinates 
$Y^M$ and the associated
momentum  $\PP_M$, i.e. 
corresponding to the Lagrangian \rf{actki}
are 
\be
{\cal L} 
= \PP_\al\dot{x}_\al 
+ \PP_M \dot{Y}^M
+\frac{{\rm i}}{2}p^+(\theta^i \dot{\theta}_i
+\eta^i\dot{\eta}_i+\theta_i \dot{\theta}^i+\eta_i\dot{\eta}^i)  -
{\cal H}  \  , 
\la{laah}
\ee
$$
{\cal H}  =  
\frac{1}{2p^+}\Bigl(\PP_\al^2 + Y^4\x'_\al^2 + 
 Y^4\PP_M\PP_M  + 
\Y'^M\Y'^M $$ \be  
+\ Y^2[p^{+2}(\eta^2)^2 + 2{\rm i}p^+\eta \rho^{MN}\eta  Y_M
\PP_N]\Bigr)
-\ |Y| Y^M \Bigl[\ \eta \rho^M 
(\th' - {\rm i}\sqrt{2}|Y| \eta\x')+h.c.\Bigr] \  , 
\la{haha}
\ee
\be
p^+\x'^-  + \PP_\al\x'_\al 
+ \PP_M \Y'^M
+\frac{{\rm i}}{2}p^+(
\theta^i\th'_i+\eta^i\et'_i
+\theta_i\th'^i+\eta_i\et'^i) =0   \ .  \la{ons}
\ee
The fact that in  the particle theory limit 
 the string Hamiltonian 
\rf{dur} or \rf{haha} reduces \ci{MTT} to the 
superparticle  Hamiltonian \ci{met3} {implies} 
that the ``massless" (zero-mode)  spectrum of the superstring
coincides indeed with the spectrum of type IIB supergravity
compactified on $S^5$. The  vertex operators for the 
supergravity states are then  obtained by solving the
linearized 
supergravity equations (or 1-st quantized superparticle state
equations) in \adss background.

To restore the dependence on $R$ and $\a'$ one needs to make
the replacements: $e^\p \to e^{\p/R}$, \ $u^M \to R u^M$, 
$\PP \to T^{-1/2} \PP$, $ x' \to  T^{1/2} x'$, etc., 
where $T= (2 \pi \a')^{-1}$. It may be interesting to 
study the limiting cases of the parameters,  like 
${\a'/ R^2}\to \infty$ for fixed momenta, etc. 
$R\to \infty$ gives of course the standard flat space 
\lc superstring Hamiltonian (in the  specific parametrization
 of fermions we are
using, see section 2). 

For generic values of parameters the 
 Hamiltonian \rf{dur},\rf{haha} 
looks quite  non-linear.
Further progress  may depend on a possibility 
of  making a transformation to some  new variables 
which  may allow one to solve  for the string theory spectrum.
Such transformation should involve  fermions 
in an essential way. The phase space formulation seems 
a good starting point for searching  for such a transformation 
since the bosonic momenta are already nontrivially dependent on
the fermions.
One potentially interesting idea is to try to introduce
 twistor-like  variables
(see  \ci{kaltw} for some previous discussions of twistors
 in $AdS$ space).

\nonumsection{Acknowledgements}
\noindent
This  work was  supported  in part by
the DOE grant DE-FG02-91ER-40690 and   the INTAS project 991590.
I am grateful to R.R. Metsaev and C.B. Thorn for  
 collaboration
on the work described in this contribution.

\nonumsection{References}
\noindent

\end{document}